\documentclass[letterpaper, 10 pt, conference]{ieeeconf}  

\IEEEoverridecommandlockouts                              
\usepackage[left=0.8in,right=0.8in,top=0.7in,bottom=0.7in]{geometry}

\usepackage{graphicx} 
\usepackage{draftwatermark}
\SetWatermarkText{DRAFT}
\SetWatermarkScale{1}

\title{\LARGE \bf
Fast Data Management with Distributed Streaming SQL
}

\author{Milinda Pathirage and Beth Plale \\
School of Informatics and Computing \\
Indiana University \\
Email: \textit{mpathira@indiana.edu}, \textit{plale@indiana.edu}}

\begin{document}

\maketitle
\thispagestyle{empty}
\pagestyle{empty}

\begin{abstract}
To stay competitive in today's data driven economy, enterprises large and small are turning to stream processing platforms  to process high volume, high velocity, and diverse streams of data (fast data) as they arrive. Low-level programming models provided by the popular systems of today suffer from lack of responsiveness to change: enhancements require code changes with attendant large turn-around times. Even though distributed SQL query engines have been available for \textit{Big Data}, we still lack support for SQL-based stream querying capabilities in distributed stream processing systems. In this white paper, we identify a set of requirements and propose a standard SQL based streaming query model for management of what has been referred to as \textit{Fast Data}.
\end{abstract}

\section{INTRODUCTION}

\textit{Big Data} can be characterized as historical data that has been processed for storage before being analyzed for insight into how users behave, organizations perform, devices or people interact. But as the data-driven economy evolves, enterprises have begun to recognize the importance of processing data as it arrives, as instant sense and response, resulting a new class of distributed processing systems known as distributed stream processing systems. The data, generated as a result of people, devices, and/or software services interacting in real time, generating massive amounts of high velocity data, is often referred to as \textit{Fast Data}.

  While earlier work on processing real time streams of data ~\cite{plale_dynamic_2003, arasu_stream:_2004, abadi_design_2005, chen_gates:_2004} focuses on high velocity data that arrives continuously (as streams of indefinite duration), the data was often of one type, for instance, finance data \cite{abadi_design_2005} or sensor data from cooperative-task robots~\cite{schroeder_software_1997}, and the solutions were often centralized. With the arrival of social media, online shopping, hand-held devices, smart sensors, health devices, and Internet of Things (IoT), the sources of information over which real time processing can be done is significantly multiplied and varied. In response, robust and scalable distributed message queue and stream processing systems like \textit{Apache Kafka}, \textit{Apache S4}, \textit{Apache Storm} and \textit{Apache Samza} have arisen to cope with the challenge. These systems allow parallel and distributed processing over numerous, heterogeneous data streams that act on data as it arrives, generating new value to both enterprises and its consumers.

  Often real-time or near real-time processing applications are backed by computed summaries or modeled information generated by traditional batch-oriented processing systems.  Too, Lambda Architecture (LA)~\cite{_lambda_}, ~\cite{boykin_summingbird:_2014}, ~\cite{yan_radstack} has seen considerable uptake as a hybrid solution to data analysis.  Architected as three layers, Lambda Architecture has a 1) speed layer for fast data, 2) batch layer for historical data, and 3) serving layer supporting queries on top of the results of the two layers. Some have gone beyond this hybrid architecture, proposing to ~\cite{kreps_questioning_} replace LA with a single distributed stream processing system backed by a message queue that supports replays and gives the scalability needed for high volume data.

Both Lambda Architecture and its follow-on suffer in that both require developers to either develop new functionality in languages like Java or Scala or use a specialized, custom query language. For instance, frameworks like Summingbird ~\cite{boykin_summingbird:_2014}, require a highly skilled developer who is able to manipulate complex programming abstractions to fully utilize the framework.  Between the high learning curve, need for skilled talent, and the too-long a turn-around time for new functionality, companies are rejecting LA as too burdensome in a world where companies have to race to outpace changes in user expectation and outperform in a fierce competitive environment.

 It has been shown through wide adoption of frameworks like \textit{Hive}, \textit{Drill} and \textit{Presto} that declarative query language access to large-scale data is a useful thing; the wide adoption alone is a strong indicator of the ease of use of SQL. In this white paper, we identify a set of requirements and propose several extensions to standard SQL for querying data streams. We also present an architecture building on the Apache incubator project "Calcite", a query planning framework that has a level of generality that makes it suitable for different query planning rules for various execution backends. We  are currently working on an implementation of the proposed query model on top Apache Samza~\cite{_high-level_}.

  \section{Requirements and SQL Extensions}
 Based on recent case studies in data stream processing applications from social media companies like LinkedIn, from recent publications from companies like Google, and through numerous discussions in open source mailing lists, we identify the following architectural and language requirements:

\textit{Architectural requirements}:

  \begin{itemize}
  \item Horizontal scalability to scale stream processing pipelines on thousands of stream partitions
  \item Fault tolerance and ability to recover by replaying checkpointed streams
  \item Out of order event handling for stream aggregations and joins
  \item Declarative processing implementable over multiple stream processing architectures
  \item Incremental processing and early results
  \end{itemize}

\textit{Language requirements}:   We advocate for the use of standard SQL with a set of minor extensions to enable streaming queries because of the wide knowledge of SQL (taught in every undergraduate computer science database course). We identified several important extensions to SQL for querying streams to meet. We note that not all standard SQL queries are possible on streams due to the blocking nature of some queries.

  \begin{itemize}
  \item Declarative query language to reduce the turnaround time for streaming application development and make it easy for non tech-savy users to develop streaming analytics tasks.
  \item Streams and relations supported as first class entities in the language and runtime. \texttt{STREAM} keyword is the main extension: tells the system to process incoming tuples, not existing ones (e.g. \texttt{SELECT STREAM * FROM OrdersStream})
  \item Window operator support preserved (hopping, tumbling and sliding).   \texttt{HOP}, \texttt{TUMBLE} functions in \texttt{GROUP BY} clause to implement hopping and tumbling window queries on stream ordered by timestamp or offset of tuple in a stream
  \item \texttt{OVER} clause for specification of a window over stream to extend \texttt{FROM} clause. Useful in windowed joins.
  \end{itemize}


  \section{Architecture}
\begin{figure}[h!]
  \centering
  \includegraphics[scale=0.7]{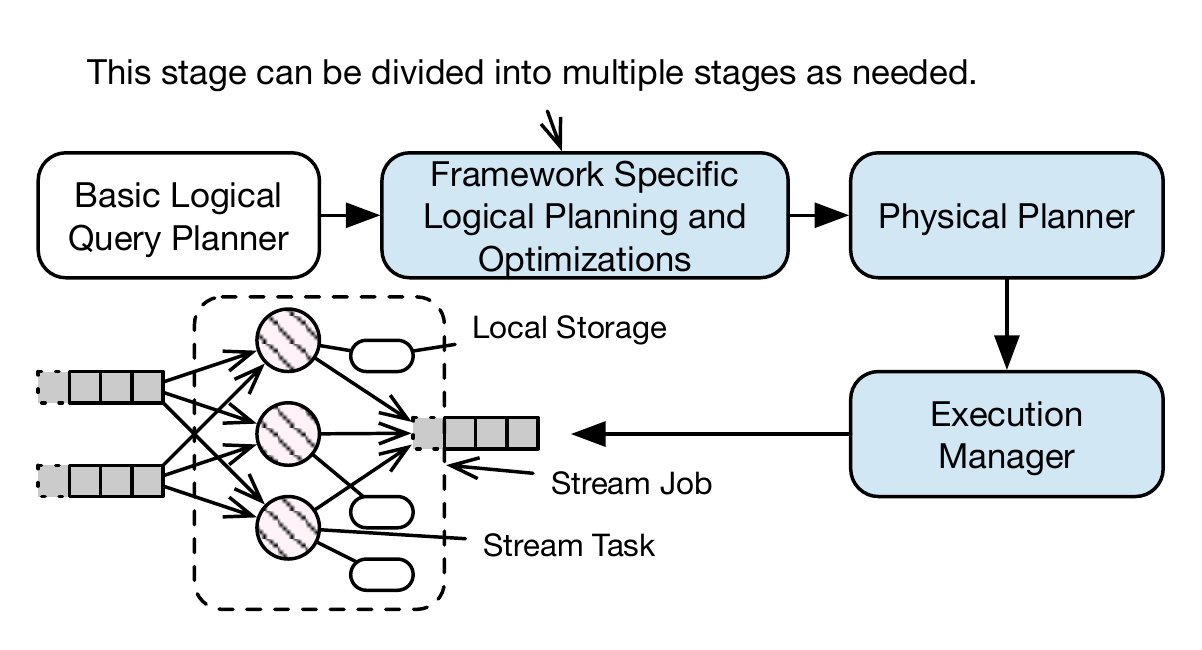}
  \caption{Proposed Architecture (Shaded stages are specific to streaming runtime)}
  \label{fig:arch}
  \end{figure}

  We propose a general architecture as a query planning pipeline, see Figure~\ref{fig:arch}, where a query is first converted into a logical plan using Apache Calcite default rules, and then several stages each implement execution back-end specific logical plan transformation rules and optimization rules.   Apache Calcite is query planning and optimization library.  The final stage transforms the logical plan to physical plan, and the physical plan is submitted as a streaming job/topology through execution manager. This architecture enables separation of logical transformations on streaming queries from the stream processing framework specific transformation and optimizations.

  Our proposed architecture best meets the requirements we laid out earlier.  Each distributed stream processing framework is different in the way it executes jobs, the type of jobs that are allowed, and the features the framework provides to the developer. Because of this variability, a query planner should consider the above variations in creating and selecting an efficient physical plan for these different runtimes. By separating out query planning into multiple stages, we allow re-use of early stages which are common to all runtimes.

  Due to varying capabilities of stream processing frameworks, achieving horizontal scalability and handling of fault tolerance and out of order arrivals will depend on the framework selection. In our case we use Apache Samza and utilize data parallelism provided by Samza based on Kafka stream partitioning to achieve horizontal scalability and stream task state is stored to checkpointed local storage for implementing fault tolerance and delayed event handling for stream aggregations and joins. We utilize Kafka's offset based message consuming architecture to implement replaying of streams by storing the last offset processed by each stream task.


  \section{Conclusion}
 A SQL-based streaming query language over fast data distributed stream processing systems will increase the turn-around times for application developers, will lower the barrier to development by allowing any developer with an understanding of SQL  to quickly adapt to the streaming settings and will make it easy to integrate relational databases into streaming queries. A general implementation of our architecture and improvements we are doing to Calcite framework as a part of this project will allow more people to extend it, re-use it and share new optimizations or planning rules in different settings.


    \section{Acknowledgement}
    \label{sec:ack}

    The authors would like to thank Julian Hyde for his contributions on streaming extensions to standard SQL and Chris Riccomini and Yi Pan for their contributions on requirements gathering, design and initial implementation of Samza SQL operator layer. Also we would like to thank Apache Samza community for their valuable feedback during initial phase of the project.

\bibliographystyle{IEEEtran}
\bibliography{streaming2015}

\end{document}